\documentclass[12pt]{article}
\usepackage{subcaption}
\usepackage{xcolor}
\usepackage{soul}
\usepackage{lineno}
\usepackage{amsmath}
\usepackage{amsfonts}
\usepackage[margin=3cm]{geometry}
\usepackage{authblk}
\usepackage{graphicx}

\title{Re-ordering of Hadamard matrix using Fourier transform and gray-level co-occurrence matrix for compressive single-pixel imaging}

\date{2022}
\author[1,*]{Pedro G. Vaz}
\author[1]{Andreia Gaudêncio}
\author[1]{L. F. Requicha Ferreira}
\author[2]{Anne Humeau-Heurtier}
\author[3,4]{Miguel Morgado}
\author[1]{João Cardoso}

\affil[1]{\small{LIBPhys-UC,%Department and Organization
            Department of Physics, Rua Larga, University of Coimbra, 
            Coimbra,
        3004-516, 
           Portugal}}
           
\affil[2]{Univ Angers, LARIS – Laboratoire Angevin de Recherche en Ingénierie des Systèmes, 62 Avenue Notre-Dame du Lac, Angers, 49000, France}   
            
\affil[3]{Department of Physics,
            Rua Larga, University of Coimbra, 
           Coimbra,
            3004-516,
            Portugal}
            
\affil[4]{CIBIT - Coimbra Institute for Biomedical Imaging and Translational Research,
Health Sciences Campus, Azinhaga de Santa Comba, 
            Coimbra,
            3000-548,
            Portugal}
          
\affil[*]{Corresponding author: pvaz@uc.pt}          
          
\begin{document}

\maketitle

\begin{abstract}
One of the most active research fields in single-pixel imaging is the influence of the sampling basis and its order in the quality of the reconstructed images. This paper presents two new orders, ascending scale (AS) and ascending inertia (AI), of the Hadamard basis and test their performance, using simulation and experimental methods, for low sampling ratios (0.5 to 0.01). These orders were compared with two state-of-the-art orders, cake-cutting (CC) and total gradient (TG), using TVAL3 as the reconstruction algorithm and three noise levels. These newly proposed orders have better reconstructed image quality on the simulation data set (110 images) and achieved structure similarity index values higher than CC order. The experimental data set (2 images) showed that the AS and AI orders performed better with a sampling ratio of 0.5, while for lower sampling ratio the performance of AS, AI and CC was similar. The TG order performed worst in the majority of the cases. Finally, the simulation results present clear evidence that peak signal-to-noise ratio (PSNR) is not a reliable image quality assessment (IQA) metric to assess image reconstruction quality in the context of single pixel imaging.
\end{abstract}


%%%%%%%%%%%%%%%%%%%%%%%%%%  body  %%%%%%%%%%%%%%%%%%%%%%%%%%

\section{Introduction}
\label{sec:introduction}

%Historical introduction
It is acknowledged that the first attempt to use a single-pixel imaging (SPI) technique may date back to the invention of the flying-spot camera during the decade of the 1920s \cite{Gray1928The}. That device consisted of a mechanical scanning imaging system aimed to transduce visual images into electrical signals, using a patterned disc with equally distanced and sized holes - the Nipkow disc. This system could be used to convert images to electrical signals in a time where two-dimensional imaging sensors were not available.

%SPI advantages
Since then, the scientific and technical improvements of this technique have transformed it into a cutting-edge technology with advantages over traditional imaging techniques. Because SPI does not require a two-dimensional sensor to produce images, it takes advantage of faster, simpler and more sensitive light detectors to produce images in low light conditions, with faster equivalent acquisition frequencies and at wavelengths outside the range of operation of two-dimensional sensors.

%SPC description
A single-pixel camera (SPC) can be constructed in two distinct configurations: selective light detection or structured illumination. In both configurations, the SPC is mainly composed of three principal components, a light source, a spatial light modulator and a photodetector. The spatial light modulator is used to selectively collect light in accordance with a set of defined patterns. The electrical signal recorded by the photodetector is then correlated with the projection/detection pattern to retrieve a reconstructed two-dimensional image.

%CS description
The reconstruction process can be combined with compressive sensing (CS), a mathematical concept that allows for signal reconstruction with a sampling frequency below the Nyquist limit. Compressive sensing in the context of SPI reduces the number of patterns needed for sampling the scene which mitigates one of its biggest disadvantages, the acquisition time.

Recently, due to the development of fast and advanced light modulators based on digital micro-mirror devices (DMD) the SPI showed advances in theoretical CS principles, like the use of new projection bases, new projection orderings, and new reconstruction algorithms, and in practical applications. Among other applications, SPI is currently used in the biomedical field for fluorescence lifetime imaging \cite{Ochoa2020High}, phosphorescence lifetime imaging \cite{Santos2021Compressive}, 3D imaging \cite{Sun2016Single} and turbid imaging \cite{Lenz2020Imaging}. All of these applications are in their early days of development and there is still room for improvements and studies to determine the best settings and image reconstruction algorithms.

This paper presents a study on the effect of the projection pattern ordering, using the Hadamard basis, in the quality of reconstructed images. Here, we present a simulation work in a large dataset (110 images), as well as an experimental work in two images, using state-of-the-art orderings, namely, the cake-cutting (CC) \cite{Yu2019Super} and total gradient (TG) \cite{Yu2020super} methods, and two novel orders: the first, denominated as ascending scale (AS), is based on the increasing of the predominant 2D spatial frequency of the projected pattern, and the second one, designated as ascending inertia (AI), is based on the increasing of inertia property of the gray-level co-occurrence matrix (GLCM) also obtained from each projected pattern. From the knowledge of the authors, both of these orders were not yet described in previous works.

\section{Theoretical overview}

\subsection{Single-pixel imaging}

As stated in the introduction, the SPI technique is able to reconstruct two-dimensional images using only a single-pixel detector. The target is illuminated with structured light produced by a DMD. The micro-mirrors are programmed to tilt in one of two directions, resulting in light projected onto the target or loss to the environment. Then, the light reflected by the target is collected by a set of optics and focused on the photodetector. 

In addition, the DMD is programmed using a sequential set of orthogonal patterns, denominated sensing basis, while the intensity detected by the photodetector is recorded using a digital acquisition system. In order to produce a complete acquisition, a number of patterns equal to the number of image pixels ($N$) should be acquired. Mathematically, this process can be translated by:

\begin{equation}
    y = \Phi x \; ,
\end{equation}

\noindent where $y \in \mathbb{R}^{N \times 1}$ is the acquired set of coefficients and $\Phi$ the sensing basis.

\subsection{Compressive sensing}

Compressive sensing allows the reconstruction of a target with a resolution of $N$ pixels by taking only $M$ measurements, being $M < N$, if two conditions are satisfied – sparsity and incoherence \cite{Duarte2008Single, Romberg2008Imaging}. The sparsity is fulfilled when a signal can be represented by high valued coefficients, in a particular orthonormal basis ($\Psi$), while the low valued coefficients are removed without significant losses. The incoherence condition implies that the mutual coherence between the sensing basis ($\Phi$) and the sparsity basis ($\Psi$) must be low to ensure a good image reconstruction. The complete mathematical description of this process can be formulated as:

\begin{equation}
    y = \Phi x = \Phi \Psi \theta = A \theta \;,
\end{equation}

\noindent where $\theta	\in \mathbb{R}^{M \times 1}$ is the set of coefficients that contains the majority of spatial information. The sampling ratio ($SR$) is defined as the ratio between the set of target measurements and the resolution of the recovered image $SR = M/N$. Some papers also denominate this quantity as compression ratio \cite{Czajkowski2018Real} but this nomenclature is counter-intuitive, and should be abandoned, because in that term ``higher compression ratios'' means the use of more sampling patterns.

\subsection{Sensing matrix and orderings}

One of the most used sensing basis is the Hadamard transform which can be represented in the form of a matrix:

\begin{equation} \label{eq:hadamard}
    H_{2^{k}} = \begin{bmatrix} 
    H_{2^{k-1}} & H_{2^{k-1}}\\
    H_{2^{k-1}} & -H_{2^{k-1}}
    \end{bmatrix} = H_2 \otimes H_{2^{k-1}} \; ,
\end{equation}

\noindent where $\otimes$ is the Kronecker product, $H_1 = 1$ and $2^k$ the number of lines of the Hadamard matrix (order). Each line of this matrix, denominated as Walsh function, is reshaped, column by column, into a 2D array when applied in SPI. One of the main characteristics of the Walsh functions is that they are only composed of two values, +1 and -1, making them appropriate to use with DMD.

The equation \ref{eq:hadamard} produces a Hadamard matrix in the natural order. Nevertheless, the order of the Walsh functions can be rearranged to allow the acquisition of higher coefficients ($\theta$) first. Many different orders have been proposed for the Hadamard matrix because the order in which the patterns are displayed has a great impact on the minimum sampling ratio that can be achieved without losing too much image quality. Here we propose two novel orders (AS and AI) and compare it with the high performing orders used in past works (CC \cite{vaz2020image} and TG \cite{Yu2020Deep}).

\subsection{Orderings}

Regarding CC, the rows of the Hadamard matrix are rearranged to increase the groups of pixels with the same value (blocks). Since this is a straightforward approach, the process will not be detailed here but it can be found in \cite{Yu2019Super, vaz2020image}.

Concerning the TG order, it was first introduced in \cite{Yu2020super} along with another order denominated as total variance ascending order. The process of determining the TG order starts by reshaping each row of the Hadamard matrix into its corresponding 2D pattern. Then, the gradient of this matrix is determined in both $x$ and $y$ direction ($G^x$ and $G^y$). Both in the original paper \cite{Yu2020super} and in our work, the Matlab function \texttt{gradient} was used to determine the mask's gradient. This function implements the central difference for the interior data and the single-sided differences along the matrix borders.

In order to archive the same order as the one presented in figure 2 of paper \cite{Yu2020super}, the equation originally presented (Eq. 5 of \cite{Yu2020super}) must be modified to:

\begin{equation}
    TG_i = \sum_{j=1}^N |G^x_j| + |G^y_j|
\end{equation}

\noindent where $TG_i$ is the total gradient of the pattern derived from Hadamard matrix line $i$, $G_j$ is the gradient value in pixel $j$ in that pattern and $N$ the number of pixels of the pattern. During this work we also tried to implement the total variation order but we were unable to reproduce the order of figure 2 of the paper \cite{Yu2020super} using the given definition (equation 3 of paper \cite{Yu2020super}).

\begin{figure}[t]
    \centering
    \includegraphics[width=\textwidth]{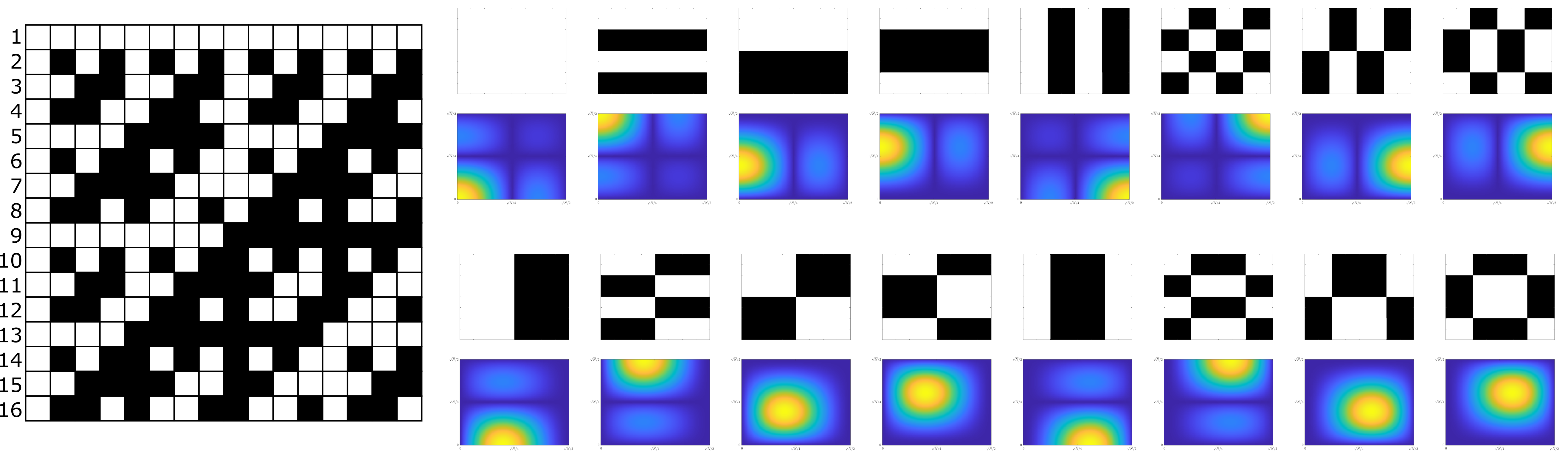}
    \caption{Hadamard matrix of order 16 in natural with the corresponding reshape patterns and respective Fourier transform first quadrant magnitude. In the 2D spectrum the DC value corresponds to the bottom left corner. Lighter color represent higher intensity.}
    \label{fig:hadamard}
\end{figure}

The newly proposed AS order is based on the identification of the predominant frequency of each reshaped Walsh function. The predominant frequency of each pattern is determined by computing the 2D Fourier transform and calculating the euclidean distance from the peak to the transform origin (0,0). Then, the Walsh functions are ordered by increasing frequency to produce the AS ordered Hadamard matrix. This process is exemplified in figure \ref{fig:hadamard}. Each line of the original Hadamard matrix is reshaped, column by column, into a 2D square pattern. From left to right and top to bottom, we can see all the 16 patterns derived from the original matrix. The magnitude of the Fourier transform first quadrant, using zero padding to 256 $\times$ 256 to increase the transform resolution, for each pattern is presented below the respective pattern, where lighter colors mean higher intensity. Finally, the patterns are ordered by ascending distance.

The proposed AI order technique is based on the inertia, also denominated contrast, property of the GLCM obtained from the reshaped Walsh function \cite{Haralick1973Textural}. First, the pattern values $-1$ are replaced by zeros in order to obtain a GLCM of the same size of the mask. Then, for each pattern we obtain the corresponding GLCMs, using 4 different directions (\mbox{$\theta$} equals to 0, 45, 90, and 135 degrees) and an offset equal to one.

Each GLCM, $g_\theta(j,k)$, will be composed of the number of times the pair of pixels $(j,k)$ occurring within the pattern. Finally, the inertia final value, for each pattern, is determined as the mean value of the 4 inertia values:

\begin{equation}
    AI_i = \frac{1}{4} \sum_{\theta=1}^{4} \sum_{j = 1}^{\sqrt{N}} \sum_{k = 1}^{\sqrt{N}} |j-k|^2 \dfrac{g_\theta(j,k)}{N_{\theta}}
\end{equation}

\noindent where $N_\theta$ is the sum of all the values of $g_\theta$

\begin{figure}[t]
\centering
\begin{subfigure}{0.24\linewidth}
\centering
\includegraphics[width=3cm]{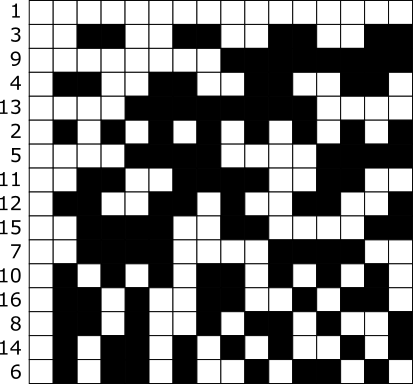}  
\caption{Cake-cutting.}
\end{subfigure}
\begin{subfigure}{0.24\linewidth}
\centering
\includegraphics[width=3cm]{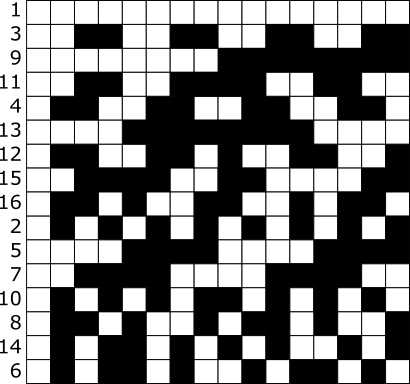}  
\caption{Ascending scale.}
\end{subfigure}
\begin{subfigure}{0.24\linewidth}
\centering
\includegraphics[width=3cm]{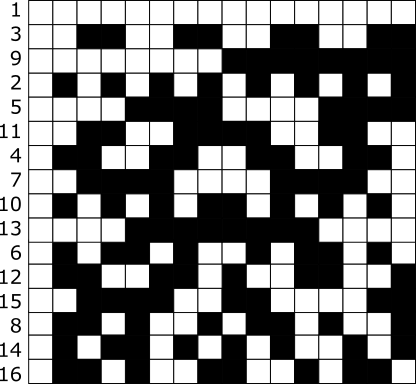}  
\caption{Total gradient.}
\end{subfigure}
\begin{subfigure}{0.24\linewidth}
\centering
\includegraphics[width=3cm]{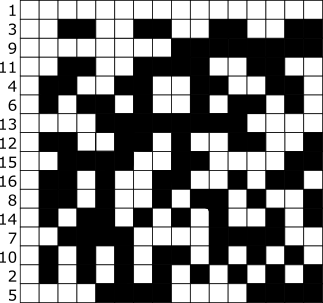}  
\caption{Ascending inertia.}
\end{subfigure}
\caption{Re-ordered Hadamard matrix (16 $\times$ 16) for different orders. The lateral numbers indicate the original position on the Natural order Hadamard matrix.}
\label{fig:matrices}
\end{figure}

Figure \ref{fig:matrices} presents the Hadamard matrices reordered for the cake-cutting (a), ascending scale (b), total gradient (c), and ascending inertia (d) orders. Most Walsh functions retain the same position. However, there are specific functions (e.g. \#16) that are substantially moved up in the AS order due to the reasons detailed above.

Although several authors agree that the most significant information in natural images is presented in the low frequency region \cite{Yu2020super}, both CC and TG orders penalize the 2D patterns that present diagonal lines, causing these patterns to be acquired last. For an illustrative example of this phenomenon, consider the Hadamard pattern \#11 (Figure \ref{fig:hadamard}), which corresponds to a diagonal low frequency stripe. Both in CC and TV orders, this pattern is selected only after patterns \#2 and \#5 which show a higher spatial frequency but in a single direction. The proposed ascending scale order aims to correct this issue, by selecting the low frequency patterns first, regardless of the stripes' direction. 

For AI order, the pattern \#11 is also selected before the patterns \#2 and \#5. The GLCMs provide the joint probability distributions of two pixel pairs, and the inertia will reflect the local variations within the matrix. Basically, the AI order allows to quantify the intensity contrast between a pixel and its neighbor along 4 directions for a given pattern. For example, both AI and AS novel orders privilege the patterns \#1, \#3, \#9, \#11, and \#4 first. However, given the higher local variations of the pattern \#6, the AI order considers it more important compared to the AS order that places it last.

\subsection{Reconstruction algorithms}

The reconstruction algorithm used during this experiment was the total variation minimization by augmented Lagrangian and alternating direction algorithm (TVAL3) \cite{Li2013efficient}. This algorithm is fast when compared to others, and finds the optimal solution by considering a sparsifying basis ($\Phi$) as the gradient of the signal by solving the equation:

\begin{equation}  \label{eq:TVAL3}
\min_{w_i x} \sum_i \|w_i\|_2, \mbox{ subject to } \Phi x = y \mbox{ and } D_i = w_i \forall_i
\end{equation}

\noindent where  $\|...\|_2$ is the $l2$ norm, $w_i = D_ix \in \mathbb{R}^{2 \times 1}$ is the discrete gradient of $x$ at position $i$, horizontally and vertically.

\subsection{Acquisition procedures}

\begin{figure}[t]
\centering
\begin{subfigure}{0.3\linewidth}
\centering
\includegraphics[width=4cm]{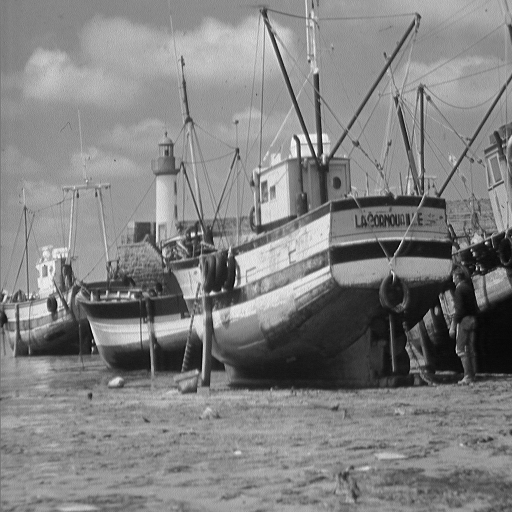}  
\caption{Boat.}
\end{subfigure}
\begin{subfigure}{0.3\linewidth}
\centering
\includegraphics[width=4cm]{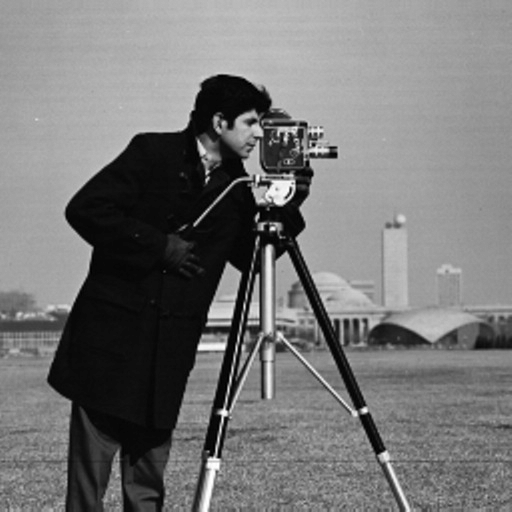}  
\caption{Cameraman.}
\end{subfigure}
\caption{Images used in the simulation and bench experiment.}
\label{fig:data-set}
\end{figure}

In this work, a dataset composed of 110 images (simulation) and two images (experimental) were used to test the reconstruction algorithm using different sampling ratios, different orderings and different noise levels for the simulation case and a data set composed of two images. The simulation dataset was composed by the MATLAB image processing toolbox built-in images. This dataset comprise a miscellaneous of natural images, including the images with different conditions of luminosity. Since the images present variations in terms of resolution and color space, all the data set was resized to a resolution of 128~$\times$~128 pixels and converted to gray scale. Figure \ref{fig:data-set} presents the two images used in both the simulation and experimental datasets, represented in gray scale levels with a resolution of 128~$\times$~128 pixels (simulation) and a size of 5~cm~$\times$~5~cm (physical paper target).

The simulated image sampling procedure was performed according to equation \eqref{eq:hadamard} where $x$ is a vectorized version of the image and $\Phi$ is the desired sensing matrix. Furthermore, normally distributed noise was added to the acquisition according to the following equation:

\begin{equation}
    y_s = y + c \times \bar{|y|} \times \sigma
\end{equation}

\noindent where $y$ represents the simulated signal projections, $c$ is a constant with values of 0 (no noise), 0.1 and 0.5, $|\bar{y}|$ is the mean of the absolute value of all projections, and $\sigma$ is an independent and normally distributed random variable with zero mean and unit standard deviation. Since this noise has a random component, five runs for each case were simulated to improve the significance of the results. 

The experimental set-up was based on a previously described single-pixel camera \cite{vaz2020image}, which uses a DMD to project the Hadamard masks into a static target. The resolution of the reconstructed images in the experimental case was 64 $\times 64$ pixels.

\subsection{Evaluation metrics}

Two distinct metrics were used to assess the simulation results: the structural similarity index (SSIM) \cite{Zhou2004Image} and the peak-signal-to-noise ratio (PSNR) \cite{Gibson2020Single}.

%Three distinct metrics were used to asses the simulation results, the image contrast \cite{Vaz2016}, the structural similarity index (SSIM) \cite{Zhou2004Image} and the peak-signal-to-noise ratio (PSNR) \cite{Gibson2020Single}. The contrast can be computed as:

%\begin{equation}
%    K = \dfrac{\sigma_{\hat{I}}}{\langle \hat{I} \rangle} \: ,
%\end{equation}

%\noindent where $\sigma_{\hat{I}}$ is the standard deviation of the reconstructed image and $\langle \hat{I} \rangle$ its mean.

The SSIM metric is focused on three image parameters: luminance ($l$), contrast ($c$) and structure ($s$). Each of these parameters is evaluated using the following Eqs.:

\begin{equation} \label{eq:lcs}
\begin{aligned}
    l(\hat{I},I) &= \frac{2\mu_{\hat{I}} \mu_I + C_1}{\mu_{\hat{I}}^2 + \mu_I^2+ C_1}   \: ,\\
    c(\hat{I},I) &= \frac{2\sigma_{\hat{I}} \sigma_I + C_2}{\sigma_{\hat{I}}^2 + \sigma_I^2 + C_2}   \: ,\\
    s(\hat{I},I) &= \frac{\sigma_{\hat{I},I} + C_3}{\sigma_{\hat{I}} \sigma_I+C_3} \: ,\\
    \text{SSIM}(\hat{I},I) &=  l(\hat{I},I)^\alpha \times c(\hat{I},I)^\beta \times s(\hat{I},I)^\gamma \: ,
\end{aligned}
\end{equation}

\noindent where $\mu$ stands for the image mean value, $\sigma$ is the image standard deviation, $\sigma_{\hat{I},I}$ represents the co-variance between $\hat{I}$ and $I$ and $C_1$, $C_2$ and $C_3$ are constants used to prevent indeterminate expressions. This metric was determined using $C_1 = 2.55^2$, $C_2 = 7.65^2$, $C3 = C2/2$ and $\alpha = \beta = \gamma = 1$.

Finally the PSNR was determined using the equation:

\begin{equation}
    PSNR = 10 \times log_{10} \dfrac{255^2}{MSE}
\end{equation}

\noindent where $MSE$ corresponds to the mean square error, and 255 corresponds to the maximum intensity of a gray scale image (8 bits image).

\section{Results}

\subsection{Simulation}

\begin{figure}[p]
    \centering
    \begin{subfigure}[t]{0.32\textwidth}
        \centering
        \includegraphics[height=3.44 cm]{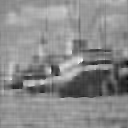}
        \caption{CC $SR=0.1$}
    \end{subfigure}%
    ~ 
    \begin{subfigure}[t]{0.32\textwidth}
        \centering
        \includegraphics[height=3.44 cm]{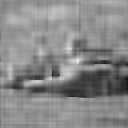}
        \caption{CC $SR=0.05$}
    \end{subfigure}
    ~
    \begin{subfigure}[t]{0.32\textwidth}
        \centering
        \includegraphics[height=3.44 cm]{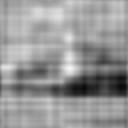}
        \caption{CC $SR=0.01$}
    \end{subfigure}
    ~
        \begin{subfigure}[t]{0.32\textwidth}
        \centering
        \includegraphics[height=3.44 cm]{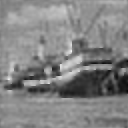}
        \caption{AS $SR=0.1$}
    \end{subfigure}%
    ~ 
    \begin{subfigure}[t]{0.32\textwidth}
        \centering
        \includegraphics[height=3.44 cm]{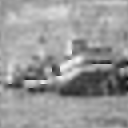}
        \caption{AS $SR=0.05$}
    \end{subfigure}
    ~
    \begin{subfigure}[t]{0.32\textwidth}
        \centering
        \includegraphics[height=3.44 cm]{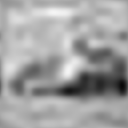}
        \caption{AS $SR=0.01$}
    \end{subfigure}
        ~
        \begin{subfigure}[t]{0.32\textwidth}
        \centering
        \includegraphics[height=3.44 cm]{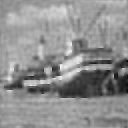}
        \caption{AI $SR=0.1$}
    \end{subfigure}%
    ~ 
    \begin{subfigure}[t]{0.32\textwidth}
        \centering
        \includegraphics[height=3.44 cm]{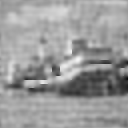}
        \caption{AI $SR=0.05$}
    \end{subfigure}
    ~
    \begin{subfigure}[t]{0.32\textwidth}
        \centering
        \includegraphics[height=3.44 cm]{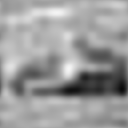}
        \caption{AI $SR=0.01$}
    \end{subfigure}
    ~
      \begin{subfigure}[t]{0.32\textwidth}
        \centering
        \includegraphics[height=3.44 cm]{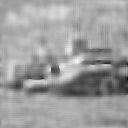}
        \caption{TG $SR=0.1$}
    \end{subfigure}%
    ~ 
    \begin{subfigure}[t]{0.32\textwidth}
        \centering
        \includegraphics[height=3.44 cm]{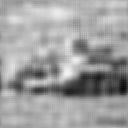}
        \caption{TG $SR=0.05$}
    \end{subfigure}
    ~
    \begin{subfigure}[t]{0.32\textwidth}
        \centering
        \includegraphics[height=3.44 cm]{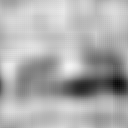}
        \caption{TG $SR=0.01$}
    \end{subfigure}
    \caption{Image reconstruction of boat using TVAL3 algorithm for the simulation with a a noise level of c = 0.1.}
    \label{fig:TVAL3_boat}
\end{figure}

\begin{figure}[p]
    \centering
    \begin{subfigure}[t]{0.32\textwidth}
        \centering
        \includegraphics[height=3.44 cm]{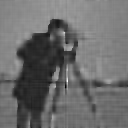}
        \caption{CC $SR = 0.1$}
    \end{subfigure}%
    ~ 
    \begin{subfigure}[t]{0.32\textwidth}
        \centering
        \includegraphics[height=3.44 cm]{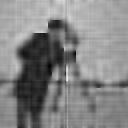}
        \caption{CC $SR = 0.05$}
    \end{subfigure}
    ~
    \begin{subfigure}[t]{0.32\textwidth}
        \centering
        \includegraphics[height=3.44 cm]{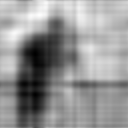}
        \caption{CC $SR = 0.01$}
    \end{subfigure}
    ~
        \begin{subfigure}[t]{0.32\textwidth}
        \centering
        \includegraphics[height=3.44 cm]{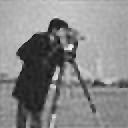}
        \caption{AS $SR = 0.1$}
    \end{subfigure}%
    ~ 
    \begin{subfigure}[t]{0.32\textwidth}
        \centering
        \includegraphics[height=3.44 cm]{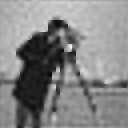}
        \caption{AS $SR = 0.05$}
    \end{subfigure}
    ~
    \begin{subfigure}[t]{0.32\textwidth}
        \centering
        \includegraphics[height=3.44 cm]{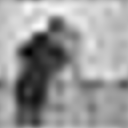}
        \caption{AS $SR = 0.01$}
    \end{subfigure}
        ~
        \begin{subfigure}[t]{0.32\textwidth}
        \centering
        \includegraphics[height=3.44 cm]{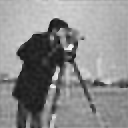}
        \caption{AI $SR = 0.1$}
    \end{subfigure}%
    ~ 
    \begin{subfigure}[t]{0.32\textwidth}
        \centering
        \includegraphics[height=3.44 cm]{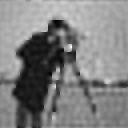}
        \caption{AI $SR = 0.05$}
    \end{subfigure}
    ~
    \begin{subfigure}[t]{0.32\textwidth}
        \centering
        \includegraphics[height=3.44 cm]{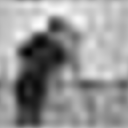}
        \caption{AI $SR = 0.01$}
    \end{subfigure}
    ~
      \begin{subfigure}[t]{0.32\textwidth}
        \centering
        \includegraphics[height=3.44 cm]{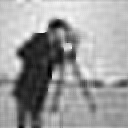}
        \caption{TG $SR = 0.1$}
    \end{subfigure}%
    ~ 
    \begin{subfigure}[t]{0.32\textwidth}
        \centering
        \includegraphics[height=3.44 cm]{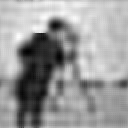}
        \caption{TG $SR = 0.05$}
    \end{subfigure}
    ~
    \begin{subfigure}[t]{0.32\textwidth}
        \centering
        \includegraphics[height=3.44 cm]{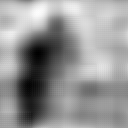}
        \caption{TG $SR = 0.01$}
    \end{subfigure}
    \caption{Image reconstruction of Cameraman using TVAL3 algorithm for the simulation with a noise level of c = 0.1.}
    \label{fig:TVAL3_cameraman}
\end{figure}

The simulation results are summarized in Figures \ref{fig:TVAL3_boat} and \ref{fig:TVAL3_cameraman}. Each figure shows the reconstructed images, of two of the used images, for the four orders, using sampling ratios of 10\%, 5\% and 1\%, and a noise level of $c = 0.1$. For all the cases, the reconstructed image quality is reduced when the sampling ratio decreases. The complete data is available at \cite{dataset}.

%Looking at the boat image, SR = 20\% case (Figure \ref{fig:TVAL3_boat}-(a), (b) and (c)), both CC and AS orders show a good image reconstruction where the shape of the boat is highly visible and the outriggers (long diagonal poles) can be clearly identified. In contrast, the image reconstructed with the TG order is blurred with less quality. The cameraman case (Figure \ref{fig:TVAL3_cameraman}-(a), (b) and (c)) presents a similar situation where CC and AS orders show similar visual quality while the TG already presents difficulties in the identification of the background, tripod and subject's face.

Looking at the boat image, when the $SR$ is equal to 10\%, significant differences between the CC and AS orders are visible (Figure \ref{fig:TVAL3_boat} - (a) and (b)). For example, the outriggers (long diagonal poles) can still be identified in the AS and AI orders while they blend in the background of the image reconstructed with CC order. The visual analysis of the images reconstructed with AS and AI show a good image reconstruction where the shape of the boat is highly visible and the outriggers can be clearly identified. In contrast, the image reconstructed with the TG order for this $SR$ are already blurred with less quality.

For the cases of $SR = 5\%$ and $SR = 1\%$, image degradation is evident in all cases. Nevertheless, the AS and AI orders are the only one able can recover the structural features with sharper edges and less blur with a SR = 5\%. Since were are dealing with low resolution images (128 $\times$ 128), none of the orders is able to recover the image structure for $SR = 1\%$.

The Cameraman case for $SR = 10\%$ (Figure \ref{fig:TVAL3_cameraman}-(a), (d), (g) and (j)) presents a similar situation where AS and AI orders show similar visual quality while the CC order already presents difficulties in the identification of the background, tripod, and camera. Nevertheless, the only order who stands out from a negative perspective is the TG order which failed to retrieve the image details. The same conclusion, with a stronger evidence, can be drawn from the Cameraman case (Figure \ref{fig:TVAL3_cameraman} - (b), (e), (h) and (k)) when the $SR$ reaches 5\%. The global image quality, specifically the background, the subject silhouette and the face details are much clearer in the AS and AI orders than in the others. Again, for the lowest $SR$ all the orders did not perform satisfactory.

\begin{figure}[p]
    \centering
    \begin{subfigure}[t]{0.49\textwidth}
        \centering
        \includegraphics[width=1\textwidth]{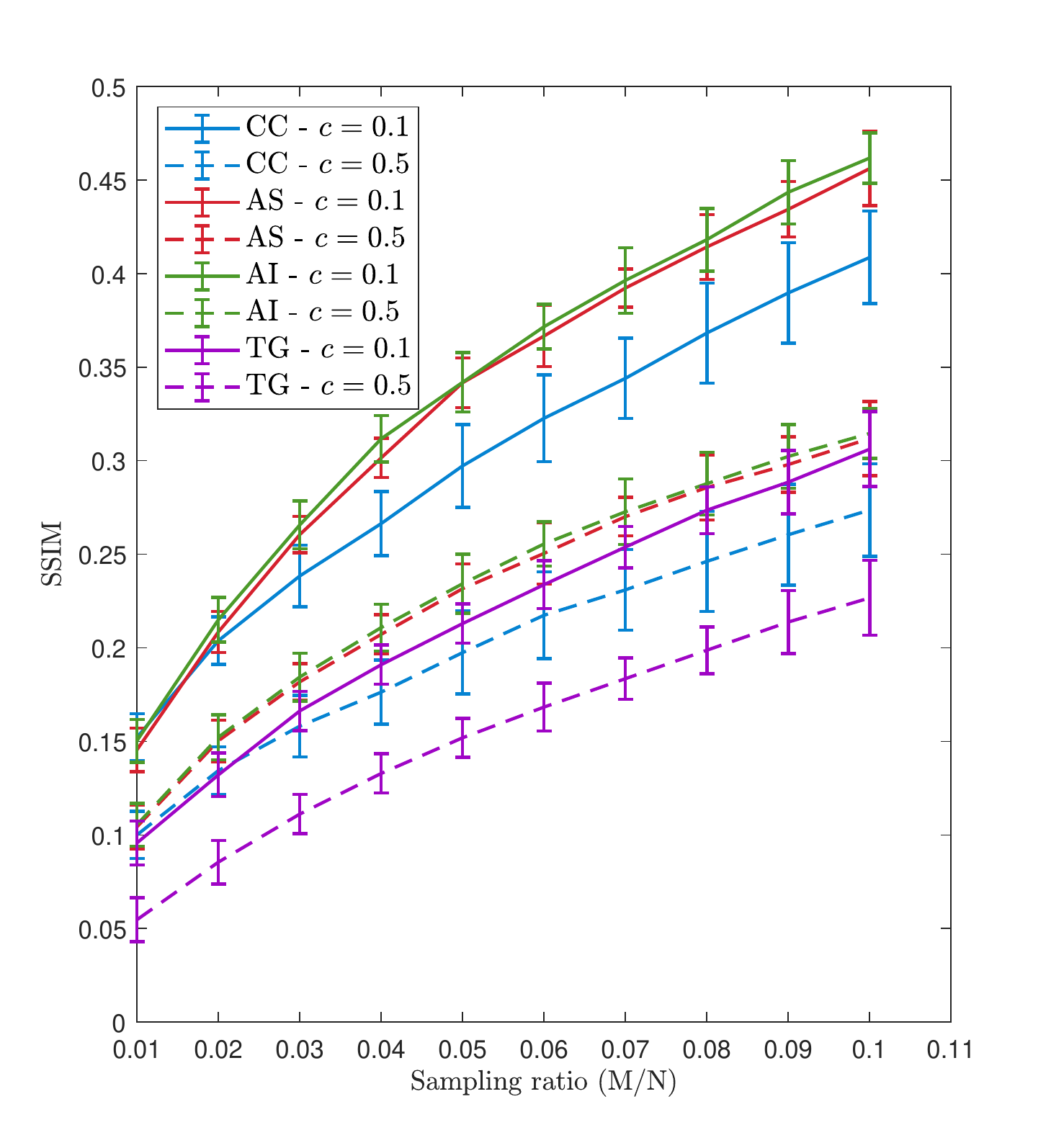}
        \caption{Dataset}
    \end{subfigure}%
    ~ 
    \begin{subfigure}[t]{0.49\textwidth}
        \centering
        \includegraphics[width=1\textwidth]{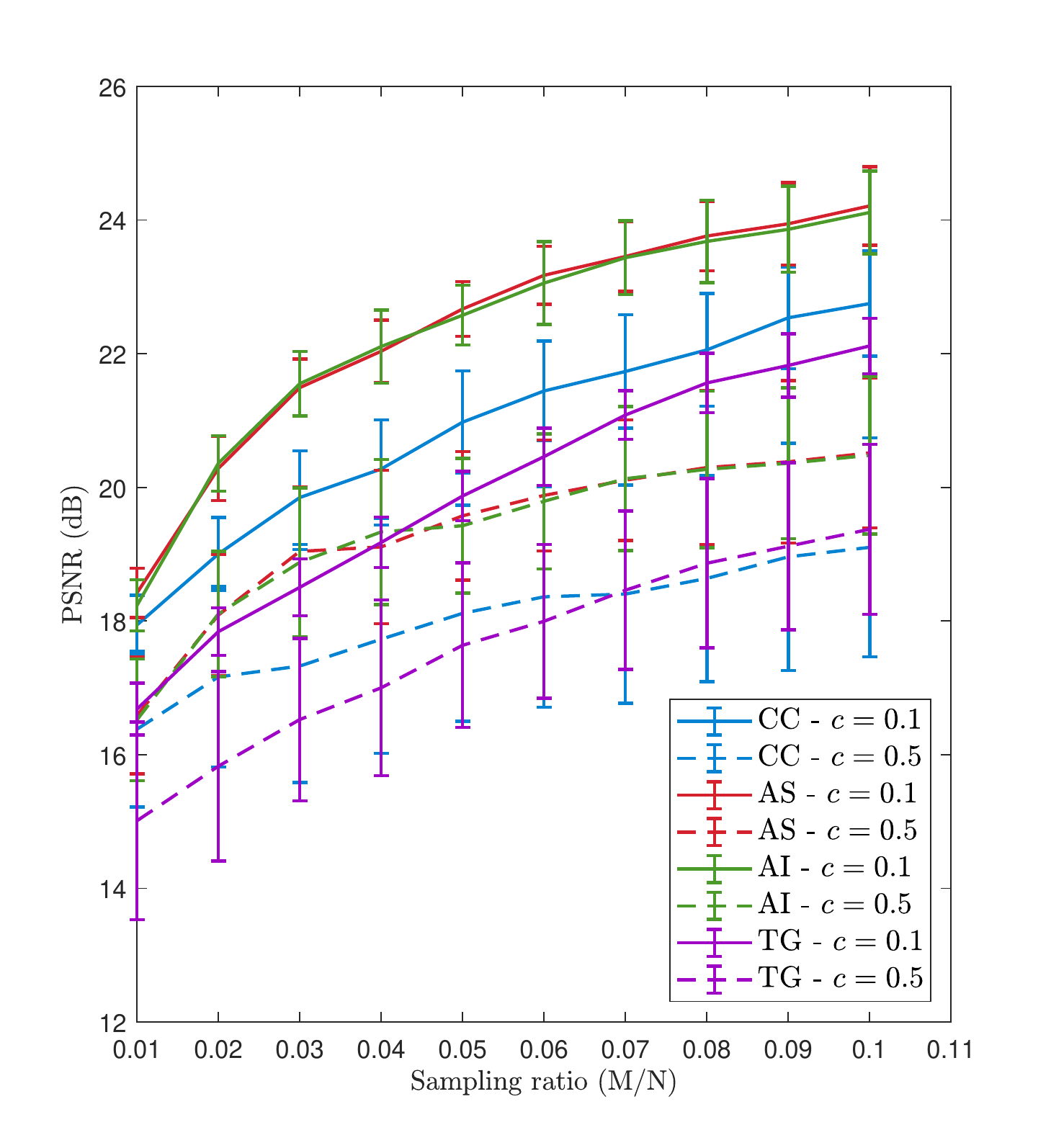}
        \caption{Dataset}
    \end{subfigure}
    ~
        \begin{subfigure}[t]{0.49\textwidth}
        \centering
        \includegraphics[width=1\textwidth]{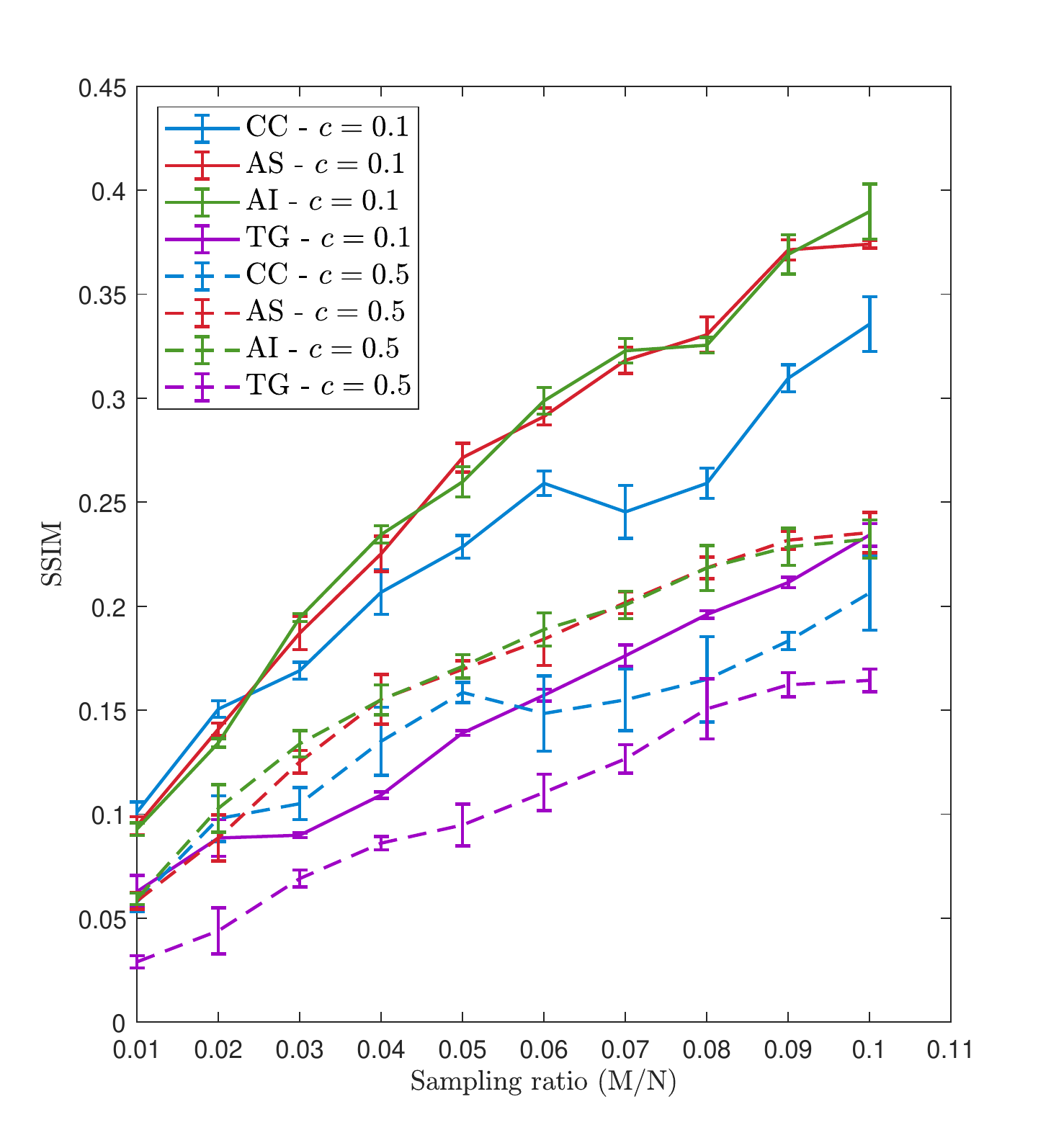}
        \caption{Cameraman}
    \end{subfigure}%
    ~ 
    \begin{subfigure}[t]{0.49\textwidth}
        \centering
        \includegraphics[width=1\textwidth]{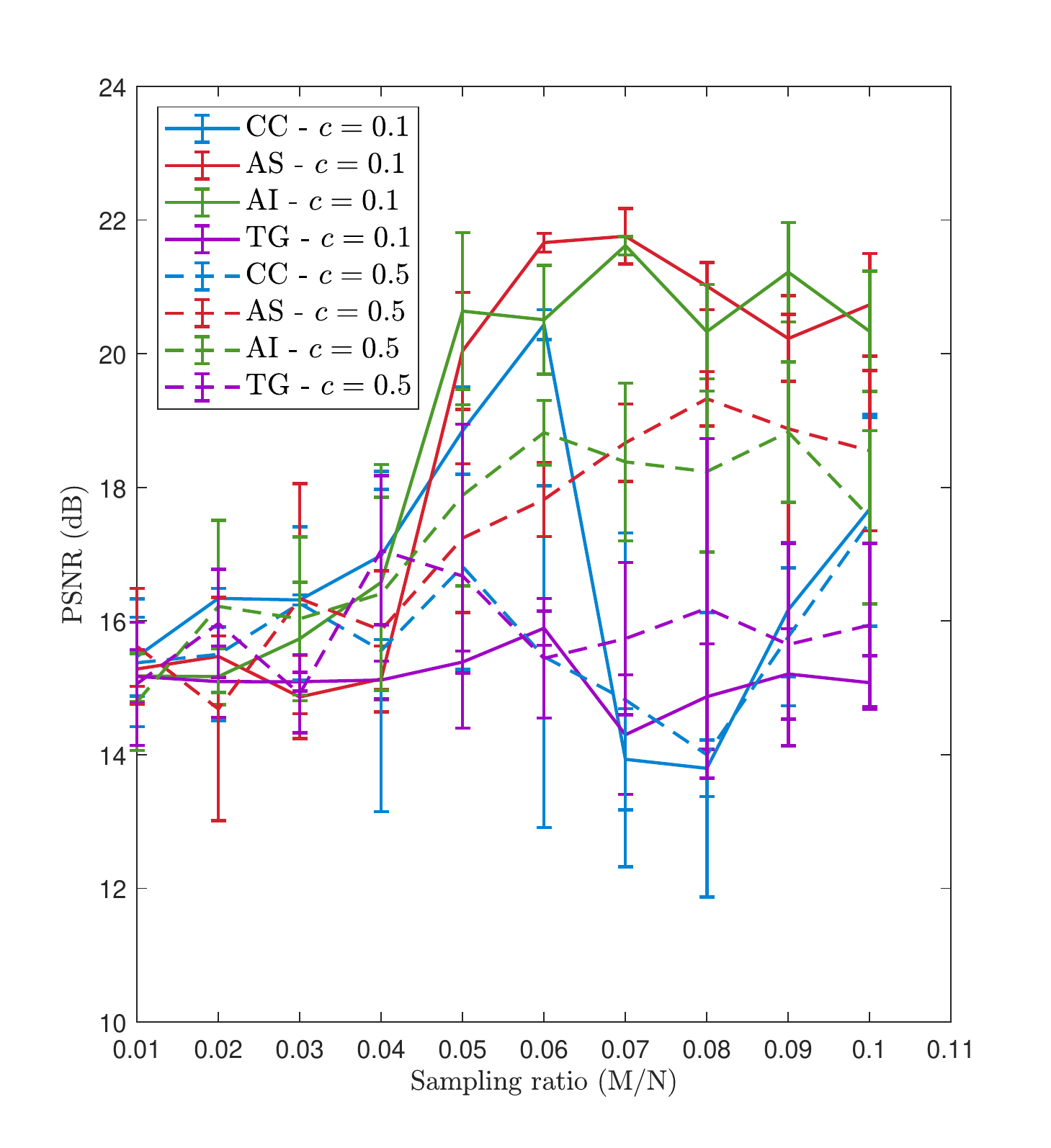}
        \caption{Cameraman}
    \end{subfigure}
    \caption{Numerical results of image reconstruction with noise levels of $c = 0.1$ and $c = 0.5$.}
    \label{fig:results_sim}
\end{figure}

The visual conclusions are also supported by the evaluation metrics (Figure \ref{fig:results_sim} - (a) and(b)). The points in the graphics correspond to the mean $\pm$ standard deviation obtained for all the dataset (110 images and 5 runs for each image). The newly proposed orders, AS and AI,  achieve better SSIM and PSNR than the state of the art TG and CC orders. An interesting phenomenon occurs when directly comparing the SSIM of AS and AI orders with CC order in the case of $c=0.1$. For sampling ratios between 0.04 and 0.1 the AS and AI orders consistently show higher SSIM when compared with the CC order. This fact is clearly reflected by the visual analysis of figures \ref{fig:TVAL3_boat} and \ref{fig:TVAL3_cameraman}. For lower $SR$ (0.01 to 0.03), the SSIM of CC order is similar the the one for AS and AI. Nevertheless, as can be seen by the reconstructed images, with these SRs most of the image structure is already lost and the SSIM could be ineffective in determining the image global quality. The influence of the noise level is also evident in the SSIM and PSNR values, causing their value to lower when the noise increase from $c=0.1$ to $c=0.5$. For the SSIM, the orders remain in the same position in terms of performance regardless of the noise level. Nevertheless, the PSNR presents a different situation, where the CC ordering is surpassed by the TG ordering for $SR>0.7$ in the noisy situation, in opposition with the less noisy situation where CC ordering is always better than TG.

If we look at the specific example of the Cameraman image (Figure \ref{fig:results_sim} - (c) and (d)), the results of PSNR show a large variation, with a non-monotonic trend, along the $SR$. The AS and AI orders still present higher PSNR, when compared with TG and CC, for most of the SRs. Moreover, the influence of the added noise on PSNR is much larger than in the SSIM as can be seen by the size of the error bars in figure when comparing figure \ref{fig:results_sim}-(c) and figure \ref{fig:results_sim}-(d) graphics.

\subsection{Experimental bench}

A global decrease in the quality of the reconstructed images can be observed when comparing these images with the simulated ones. The signal-to-noise ratio in a real environment is affected by many different noise sources that degrade the signal quality. Among those noise sources, the large noise equivalent power of the photodiode \cite{vaz2020image} and the projection system focus that slightly blurs the pattern on the test image must be highlighted.

\begin{figure}[p]
    \centering
    \begin{subfigure}[t]{0.32\textwidth}
        \centering
        \includegraphics[height=3.44 cm]{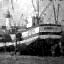}
        \caption{CC $SR = 0.5$}
    \end{subfigure}%
    ~ 
    \begin{subfigure}[t]{0.32\textwidth}
        \centering
        \includegraphics[height=3.44 cm]{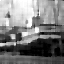}
        \caption{CC $SR = 0.2$}
    \end{subfigure}
    ~
    \begin{subfigure}[t]{0.32\textwidth}
        \centering
        \includegraphics[height=3.44 cm]{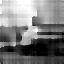}
        \caption{CC $SR = 0.05$}
    \end{subfigure}
    ~
    \begin{subfigure}[t]{0.32\textwidth}
        \centering
        \includegraphics[height=3.44 cm]{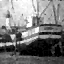}
        \caption{AS $SR = 0.5$}
    \end{subfigure}%
    ~ 
    \begin{subfigure}[t]{0.32\textwidth}
        \centering
        \includegraphics[height=3.44 cm]{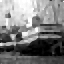}
        \caption{AS $SR = 0.2$}
    \end{subfigure}
    ~
    \begin{subfigure}[t]{0.32\textwidth}
        \centering
        \includegraphics[height=3.44 cm]{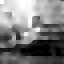}
        \caption{AS $SR = 0.05$}
    \end{subfigure}
        ~
        \begin{subfigure}[t]{0.32\textwidth}
            \centering
            \includegraphics[height=3.44 cm]{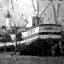}
            \caption{AI $SR = 0.5$}
        \end{subfigure}%
        ~ 
        \begin{subfigure}[t]{0.32\textwidth}
            \centering
            \includegraphics[height=3.44 cm]{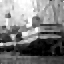}
            \caption{AI $SR = 0.2$}
        \end{subfigure}
        ~
        \begin{subfigure}[t]{0.32\textwidth}
            \centering
            \includegraphics[height=3.44 cm]{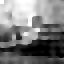}
            \caption{AI $SR = 0.05$}
        \end{subfigure}
    ~
    \begin{subfigure}[t]{0.32\textwidth}
        \centering
        \includegraphics[height=3.44 cm]{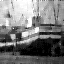}
        \caption{TG $SR = 0.5$}
    \end{subfigure}%
    ~ 
    \begin{subfigure}[t]{0.32\textwidth}
        \centering
        \includegraphics[height=3.44 cm]{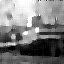}
        \caption{TG $SR = 0.2$}
    \end{subfigure}
    ~
    \begin{subfigure}[t]{0.32\textwidth}
        \centering
        \includegraphics[height=3.44 cm]{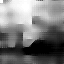}
        \caption{TG $SR = 0.05$}
    \end{subfigure}
    \caption{Experimental image reconstruction of boat.}
    \label{fig:TVAL3_boat_exp}
\end{figure}

Due to the lower image reconstruction quality of the experimental bench data, the presented sampling ratio is higher than the one used for the simulation case. When looking at the results of boat image reconstruction (figure \ref{fig:TVAL3_boat_exp}) for $SR=0.5$, the differences between the four orders are not as evident as the differences presented in the simulation results. For lower sampling ratios ($SR \leq 0.2$), CC, AS and AI have a better reconstruction image quality than TG. This is in accordance with the simulation results where AS and AI presented very similar performances, followed by CC, while TG was by far the ordering with the worst reconstruction performance (considering SSIM and PSNR). 

\begin{figure}[p]
    \centering
    \begin{subfigure}[t]{0.32\textwidth}
        \centering
        \includegraphics[height=3.44 cm]{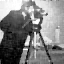}
        \caption{CC $SR = 0.5$}
    \end{subfigure}%
    ~ 
    \begin{subfigure}[t]{0.32\textwidth}
        \centering
        \includegraphics[height=3.44 cm]{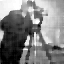}
        \caption{CC $SR = 0.2$}
    \end{subfigure}
    ~
    \begin{subfigure}[t]{0.32\textwidth}
        \centering
        \includegraphics[height=3.44 cm]{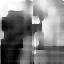}
        \caption{CC $SR = 0.05$}
    \end{subfigure}
    ~
    \begin{subfigure}[t]{0.32\textwidth}
        \centering
        \includegraphics[height=3.44 cm]{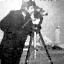}
        \caption{AS $SR = 0.5$}
    \end{subfigure}%
    ~ 
    \begin{subfigure}[t]{0.32\textwidth}
        \centering
        \includegraphics[height=3.44 cm]{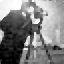}
        \caption{AS $SR = 0.2$}
    \end{subfigure}
    ~
    \begin{subfigure}[t]{0.32\textwidth}
        \centering
        \includegraphics[height=3.44 cm]{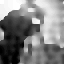}
        \caption{AS $SR = 0.05$}
    \end{subfigure}
        ~
        \begin{subfigure}[t]{0.32\textwidth}
            \centering
            \includegraphics[height=3.44 cm]{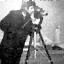}
            \caption{AI $SR = 0.5$}
        \end{subfigure}%
        ~ 
        \begin{subfigure}[t]{0.32\textwidth}
            \centering
            \includegraphics[height=3.44 cm]{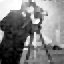}
            \caption{AI $SR = 0.2$}
        \end{subfigure}
        ~
        \begin{subfigure}[t]{0.32\textwidth}
            \centering
            \includegraphics[height=3.44 cm]{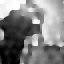}
            \caption{AI $SR = 0.05$}
        \end{subfigure}
    ~
    \begin{subfigure}[t]{0.32\textwidth}
        \centering
        \includegraphics[height=3.44 cm]{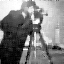}
        \caption{TG $SR = 0.5$}
    \end{subfigure}%
    ~ 
    \begin{subfigure}[t]{0.32\textwidth}
        \centering
        \includegraphics[height=3.44 cm]{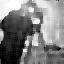}
        \caption{TG $SR = 0.2$}
    \end{subfigure}
    ~
    \begin{subfigure}[t]{0.32\textwidth}
        \centering
        \includegraphics[height=3.44 cm]{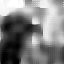}
        \caption{TG $SR = 0.05$}
    \end{subfigure}
    \caption{Experimental image reconstruction of cameraman.}
    \label{fig:TVAL3_cameraman_exp}
\end{figure}

In the case of the cameraman image (figure \ref{fig:TVAL3_cameraman_exp}), the AS and AI orders show a better definition for the case when $SR=0.5$. As an example, the subjects facial definition in the AS and AI orders is enough for its eye to be visible while it is not visible in both CC and TG order. Moreover, the tripod right leg bracing is more defined in the AS and AI orders. Nevertheless, the performance of AS, AI and CC is similar, when analyzing the images for $SR = 0.2$, while TG shows a reconstruction with lower quality. For the case when $SR = 0.05$ all the orders present lower reconstruction quality but differences can still be highlighted. TG reconstruction is more blurred when comparing with the others. When analyzing the shoulder of the cameraman, AS order provides a better reconstruction, with less aliasing. Both CC and AI reconstruct this area as a stair while the AS order provides a more reliable reconstruction with a diagonal line. This is in line with the theoretical considerations where patterns with diagonal lines are flavored in the AS order which results in a better reconstruction of these features when low SR are used.

Regarding the numerical results, the absence of an aligned ground truth image invalidate the use of SSIM and PSNR as a robust measurement metric. The PSNR method is based on pixel by pixel comparison which is extremely dependent on the correct alignment between the reconstructed image and the ground truth. On the other hand, SSIM depends on the co-variance between both images which should be determined using the exactly same field for view. When performing an experimental acquisition, it is difficult to ensure that the projected pattern illuminates the same field of view as the original target image. For example, in the presented case, the reconstructed experimental image is a cropped version of the original target. Alternative no-reference image quality assessment (NR-IQA) metrics can be used to overcame these issues. Several methods based on traditional and machine learning methods have been developed in recent years \cite{zhu2020MetaIQA, varga2021no}.

Since comparison of NR-IQA algorithms is out of the scope, we have applied three of the most used state of the art methods, namely: the blind/referenceless image spatial quality evaluator  (BRISQUE) \cite{mittal2012no}, the natural image quality evaluator (NIQE) \cite{mittal2012making} and the perception based image quality evaluator (PIQE) to our reconstructed images \cite{venkatanath2015Blind}. Nevertheless, none of the NR-IQA metrics prove to be conclusive in the evaluation of these images. The low resolution of the images and the fact that they all show artifacts from the compressive sampling reconstruction could be a possible explanation for this occurrence. Moreover this type of IQA were design to assess images with higher resolutions and are only sensitive to minimal level of noise.

\section{Discussion}
\label{sec:conclusion}

The simulation experiment indicates that the newly proposed ascending scale and ascending inertia orders achieve better image quality in the reconstruction of a diverse dataset. This conclusion is backed by the visual inspection of the images and by the SSIM and PSNR values achieved by these orders in most of the tested $SR$ range. The cake-cutting order obtains the best SSIM but only for low sampling ratios, where the image quality is already poor. In our experimental conditions, the total gradient order showed lower image reconstruction quality for all the SRs. Both visual inspection and numerical values strongly back this conclusion.

In contrast, the experimental bench results do not fully confirm this conclusion. While the ascending order and ascending inertia achieve better image quality for $SR = 0.5$, for lower sampling ratios, the difference in image reconstruction quality between cake-cutting and the remaining orders is small. In addition, it was expected for the experimental sampling ratio to be higher since low resolution SPI coefficients tend to be less sparse \cite{vaz2020image, Yu2019Super} leading to problems in the reconstruction process \cite{Sun2016Single}. 

In addition, as can be seen in figure \ref{fig:results_sim} (d), PSNR is an evaluation metric very sensitive to the image noise component. The large error bars presented for the same conditions show that very different PSNR values are obtained for images with similar visual quality. The PSNR should be used very carefully when in presence of noise sources or even small spatial displacements of the target because it is computed using point-wise computations.

As our results demonstrate, PSNR can be misleading even in low noise conditions. For the boat image case (Fig. \ref{fig:results_sim} (b) - $c=0.1$), CC order achieved a PSNR = 21 dB while AS achieved a PSNR = 19 dB for a sampling ratio of 0.05. A visual analysis of these images shows that the image reconstructed with AS order has better quality than the image reconstructed with CC order.

The discussion about which Hadamard ordering allows for the lower sampling ratio gave rise to many different orderings \cite{Yu2019Super, Sun2017A, vaz2020image, Yu2020super} in recent works. Although we can conclude that some are better than other (for example natural order vs Walsh order \cite{vaz2020image} or CC vs Paley \cite{Yu2020super}), the best order will depend on the structure of the image and on the noise conditions. Other works \cite{Yu2020Deep} have also showed that the best order can change depending on the analyzed image. For example, in \cite{Yu2020Deep}, for the reconstruction of an image of a man, the best order was the total variation while the power order achieved better results when reconstructing a wheel image. In the conditions previously documented in our work, ascending scale and ascending inertia surpassed the cake-cutting and, by a large margin, the total gradient order.

Single-pixel imaging range of applications has been extending in recent years including microscopy \cite{Peng2021Fourier}, phophorescence \cite{Santos2021Compressive} and fluorescence imaging, microtomography \cite{Peng2018Micro}, retinal imaging \cite{Dutta2019Single} and in telescopic systems \cite{Zhang2020Dual}. Each one of these applications has its own requirements and produces different images. Researchers doing single-pixel imaging should not exclude, \textit{a priori}, one or other order. Instead, simulation works should be performed using typical images with similar structure to the real ones to determine the order/orders which allow for the lowest sampling ratio. Another approach is to use an adaptive strategy as the one developed for Fourier sampling basis \cite{yuan2021adaptive}.

\section{Conclusions}

The presented paper proposes two novel Hadamard ordering techniques for SPI applications based on Fourier analysis and on the inertia property of GLCM, and are denominated as ascending scale and ascending inertia, respectively. The simulation results, using a dataset of 110 images, show that these orders produce better image quality when compared with state-of-the-art methods. The experimental results indicate that the ascending scale and ascending inertia orders have the potential to be good candidates to achieve lower sampling ratios.

\section{Funding}
Fundação para a Ciência e a Tecnologia (PTDC/EMD-TLM/30295/2017); European Regional
Development Fund; Competitiveness and Internationalization Operational Programme (PTCOMPETE 2020).

\section{Disclosures}

The authors declare no conflicts of interest.

\bibliographystyle{elsarticle-num}
\bibliography{SPI}

\begin{thebibliography}{10}
\expandafter\ifx\csname url\endcsname\relax
  \def\url#1{\texttt{#1}}\fi
\expandafter\ifx\csname urlprefix\endcsname\relax\def\urlprefix{URL }\fi
\expandafter\ifx\csname href\endcsname\relax
  \def\href#1#2{#2} \def\path#1{#1}\fi

\bibitem{Gray1928The}
F.~Gray, The use of a moving beam of light to scan a scene for television, J.
  Opt. Soc. Am. 16~(3) (1928) 177--190.
\newblock \href {https://doi.org/10.1364/JOSA.16.000177}
  {\path{doi:10.1364/JOSA.16.000177}}.

\bibitem{Ochoa2020High}
M.~Ochoa, A.~Rudkouskaya, R.~Yao, P.~Yan, M.~Barroso, X.~Intes, High
  compression deep learning based single-pixel hyperspectral macroscopic
  fluorescence lifetime imaging in vivo, Biomed. Opt. Express 11~(10) (2020)
  5401--5424.
\newblock \href {https://doi.org/10.1364/BOE.396771}
  {\path{doi:10.1364/BOE.396771}}.

\bibitem{Santos2021Compressive}
P.~P. Santos, P.~G. Vaz, A.~S.~F. Gaud{\^e}ncio, M.~Morgado, N.~A. Pereira,
  M.~Pineiro, T.~P. e~Melo, J.~Cardoso, Compressive single pixel
  phosphorescence lifetime and intensity simultaneous imaging: a pilot study
  using oxygen sensitive biomarkers, in: Integrated Optics: Design, Devices,
  Systems and Applications VI, Vol. 11775, International Society for Optics and
  Photonics, 2021, p. 1177511.

\bibitem{Sun2016Single}
M.-J. Sun, M.~P. Edgar, G.~M. Gibson, B.~Sun, N.~Radwell, R.~Lamb, M.~J.
  Padgett, Single-pixel three-dimensional imaging with time-based depth
  resolution, Nature Communications 7~(1) (jul 2016).
\newblock \href {https://doi.org/10.1038/ncomms12010}
  {\path{doi:10.1038/ncomms12010}}.

\bibitem{Lenz2020Imaging}
A.~J.~M. Lenz, P.~C. Pesudo, V.~Climent, J.~Lancis, E.~Tajahuerce, {Imaging the
  optical properties of turbid media with single-pixel detection}, in:
  C.~Fournier, M.~P. Georges, G.~Popescu (Eds.), Unconventional Optical Imaging
  II, Vol. 11351, International Society for Optics and Photonics, SPIE, 2020,
  pp. 35 -- 40.
\newblock \href {https://doi.org/10.1117/12.2554561}
  {\path{doi:10.1117/12.2554561}}.

\bibitem{Yu2019Super}
W.-K. Yu, \href{https://www.mdpi.com/1424-8220/19/19/4122}{Super sub-nyquist
  single-pixel imaging by means of cake-cutting hadamard basis sort}, Sensors
  19~(19) (2019).
\newblock \href {https://doi.org/10.3390/s19194122}
  {\path{doi:10.3390/s19194122}}.
\newline\urlprefix\url{https://www.mdpi.com/1424-8220/19/19/4122}

\bibitem{Yu2020super}
X.~Yu, R.~I. Stantchev, F.~Yang, E.~Pickwell-MacPherson, Super sub-nyquist
  single-pixel imaging by total variation ascending ordering of the hadamard
  basis, Scientific Reports 10~(1) (2020) 1--11.

\bibitem{Duarte2008Single}
M.~F. Duarte, M.~A. Davenport, D.~Takhar, J.~N. Laska, T.~Sun, K.~F. Kelly,
  R.~G. Baraniuk, Single-pixel imaging via compressive sampling, IEEE signal
  processing magazine 25~(2) (2008) 83--91.

\bibitem{Romberg2008Imaging}
J.~Romberg, Imaging via compressive sampling, IEEE Signal Processing Magazine
  25~(2) (2008) 14--20.

\bibitem{Czajkowski2018Real}
K.~M. Czajkowski, A.~Pastuszczak, R.~Koty{\'n}ski, Real-time single-pixel video
  imaging with fourier domain regularization, Optics express 26~(16) (2018)
  20009--20022.

\bibitem{vaz2020image}
P.~G. Vaz, D.~Amaral, L.~R. Ferreira, M.~Morgado, J.~Cardoso, Image quality of
  compressive single-pixel imaging using different hadamard orderings, Optics
  express 28~(8) (2020) 11666--11681.

\bibitem{Yu2020Deep}
X.~Yu, F.~Yang, B.~Gao, J.~Ran, X.~Huang, Deep compressive single pixel imaging
  by reordering hadamard basis: A comparative study, IEEE Access 8 (2020)
  55773--55784.
\newblock \href {https://doi.org/10.1109/ACCESS.2020.2981505}
  {\path{doi:10.1109/ACCESS.2020.2981505}}.

\bibitem{Haralick1973Textural}
R.~M. Haralick, K.~Shanmugam, I.~H. Dinstein, Textural features for image
  classification, IEEE Transactions on systems, man, and cybernetics~(6) (1973)
  610--621.

\bibitem{Li2013efficient}
C.~Li, W.~Yin, H.~Jiang, Y.~Zhang, An efficient augmented lagrangian method
  with applications to total variation minimization, Computational Optimization
  and Applications 56~(3) (2013) 507--530.

\bibitem{Zhou2004Image}
{Zhou Wang}, A.~C. {Bovik}, H.~R. {Sheikh}, E.~P. {Simoncelli}, Image quality
  assessment: from error visibility to structural similarity, IEEE Transactions
  on Image Processing 13~(4) (2004) 600--612.
\newblock \href {https://doi.org/10.1109/TIP.2003.819861}
  {\path{doi:10.1109/TIP.2003.819861}}.

\bibitem{Gibson2020Single}
G.~M. Gibson, S.~D. Johnson, M.~J. Padgett, Single-pixel imaging 12 years on: a
  review, Opt. Express 28~(19) (2020) 28190--28208.
\newblock \href {https://doi.org/10.1364/OE.403195}
  {\path{doi:10.1364/OE.403195}}.

\bibitem{dataset}
P.~Vaz, A.~Gaudêncio, M.~Morgado, J.~Cardoso,
  \href{https://gitlab.com/l3151/spi}{Compressive single pixel imaging
  simulation and experimental results}, Gitlab (2022).
\newline\urlprefix\url{https://gitlab.com/l3151/spi}

\bibitem{zhu2020MetaIQA}
H.~Zhu, L.~Li, J.~Wu, W.~Dong, G.~Shi, Metaiqa: Deep meta-learning for
  no-reference image quality assessment, in: Proceedings of the IEEE/CVF
  Conference on Computer Vision and Pattern Recognition, 2020, pp.
  14143--14152.

\bibitem{varga2021no}
D.~Varga, No-reference image quality assessment with multi-scale orderless
  pooling of deep features, Journal of Imaging 7~(7) (2021) 112.

\bibitem{mittal2012no}
A.~Mittal, A.~K. Moorthy, A.~C. Bovik, No-reference image quality assessment in
  the spatial domain, IEEE Transactions on image processing 21~(12) (2012)
  4695--4708.

\bibitem{mittal2012making}
A.~Mittal, R.~Soundararajan, A.~C. Bovik, Making a “completely blind” image
  quality analyzer, IEEE Signal processing letters 20~(3) (2012) 209--212.

\bibitem{venkatanath2015Blind}
N.~Venkatanath, D.~Praneeth, M.~C. Bh, S.~S. Channappayya, S.~S. Medasani,
  Blind image quality evaluation using perception based features, in: 2015
  Twenty First National Conference on Communications (NCC), IEEE, 2015, pp.
  1--6.

\bibitem{Sun2017A}
M.-J. Sun, L.-T. Meng, M.~P. Edgar, M.~J. Padgett, N.~Radwell, A russian dolls
  ordering of the hadamard basis for compressive single-pixel imaging,
  Scientific reports 7~(1) (2017) 3464.

\bibitem{Peng2021Fourier}
J.~Peng, M.~Yao, Z.~Huang, J.~Zhong, Fourier microscopy based on single-pixel
  imaging for multi-mode dynamic observations of samples, APL Photonics 6~(4)
  (2021) 046102.

\bibitem{Peng2018Micro}
J.~Peng, M.~Yao, J.~Cheng, Z.~Zhang, S.~Li, G.~Zheng, J.~Zhong,
  Micro-tomography via single-pixel imaging, Optics express 26~(24) (2018)
  31094--31105.

\bibitem{Dutta2019Single}
R.~Dutta, S.~Manzanera, A.~Gamb{\'\i}n-Regadera, E.~Irles, E.~Tajahuerce,
  J.~Lancis, P.~Artal, Single-pixel imaging of the retina through scattering
  media, Biomedical optics express 10~(8) (2019) 4159--4167.

\bibitem{Zhang2020Dual}
Y.~Zhang, G.~M. Gibson, M.~P. Edgar, G.~Hammond, M.~J. Padgett, Dual-band
  single-pixel telescope, Optics Express 28~(12) (2020) 18180--18188.

\bibitem{yuan2021adaptive}
A.~Y. Yuan, J.~Feng, S.~Jiao, Y.~Gao, Z.~Zhang, Z.~Xie, L.~Du, T.~Lei, Adaptive
  and dynamic ordering of illumination patterns with an image dictionary in
  single-pixel imaging, Optics Communications 481 (2021) 126527.

\end{thebibliography}

\vfill
\end{document}